\documentclass[%
 aip,
 amsmath,amssymb,
 reprint,%
]{revtex4-1}

\usepackage{graphicx}% Include figure files
\usepackage{dcolumn}% Align table columns on decimal point
\usepackage{bm}% bold math
\usepackage[utf8]{inputenc}
\usepackage[T1]{fontenc}
\usepackage{mathptmx}

\begin{document}

\preprint{AIP/123-QED}

\title[]{Role of self-generated magnetic fields in the inertial fusion ignition threshold}

\author{James D. Sadler}
\email{sadler6@llnl.gov}
 \affiliation{%
Lawrence Livermore National Laboratory, 7000 East Avenue, Livermore, CA 94550, USA}%
 \affiliation{%
Theoretical Division, Los Alamos National Laboratory, Los Alamos, NM 87545, USA}%
\author{Christopher A. Walsh}
 \affiliation{%
Lawrence Livermore National Laboratory, 7000 East Avenue, Livermore, CA 94550, USA}%
\author{Ye Zhou}
 \affiliation{%
Lawrence Livermore National Laboratory, 7000 East Avenue, Livermore, CA 94550, USA}%
\author{Hui Li}
 \affiliation{%
Theoretical Division, Los Alamos National Laboratory, Los Alamos, NM 87545, USA}%

\date{\today}% It is always \today, today,
\begin{abstract}
   Magnetic fields spontaneously grow at unstable interfaces around hot-spot asymmetries during inertial confinement fusion implosions. Although difficult to measure, theoretical considerations and numerical simulations predict field strengths exceeding $5\,$kT in current national ignition facility experiments. Magnetic confinement of electrons then reduces the hot-spot heat loss by $>5$\%. We demonstrate this via magnetic post-processing of two-dimensional xRAGE hydrodynamic simulation data at bang time. We then derive a model for the self-magnetization, finding that it varies with the square of the stagnation temperature and inversely with the areal density. The self-magnetized Lawson analysis then gives a slightly reduced ignition threshold. Time dependent hot-spot energy balance models corroborate this finding, with the magnetic field quadrupling the fusion yield for near threshold parameters. The inclusion of magnetized multi-dimensional fluid instabilities could further alter the ignition threshold, and will be the subject of future work.
\end{abstract}
\maketitle

\section{Introduction}

Inertial confinement fusion (ICF) uses lasers to compress a deuterium-tritium (DT) fuel capsule \cite{nuckolls1972laser, lindl1995development, zylstra2021record} to temperatures $T>4\,$keV and mass density $\rho>100\,$gcm$^{-3}$. Under these conditions, fusion burn can deposit more energy into the fuel than that which is lost via radiation, conduction and expansion. Since the thermonuclear reaction rate increases strongly with temperature, this ignites a runaway fusion burn \cite{christopherson2020theory}. A significant fraction of the fuel mass can be burned, leading to tens of MJ fusion yield from less than one milligram of fuel. Implosions aim to create a hot region with $T>4\,$keV, containing a small fraction of the fuel. It ignites the rest of the fuel if plasma conditions exceed a certain threshold, primarily dependent on the hot-spot temperature and areal density \cite{lawson1957some, hurricane2018beyond}. This can be intuitively understood by considering the strong increase of fusion rate with temperature, and also that the plasma size must be sufficient to overcome conduction and expansion losses during fusion self-heating. Further work has quantified the proximity to ignition in terms of measurable experimental quantities \cite{zhou2008measurable, betti2010thermonuclear, haan2011point}. The ignition threshold is also theoretically linked to the yield amplification, which is the ratio of the yield to that of an equivalent implosion without fusion energy deposition \cite{christopherson2020theory, hurricane2021thermodynamic}. 

A simple ignition threshold can be estimated by requiring the energy deposition of fast alpha particles from DT fusion reactions to exceed the losses to conduction and radiation \cite{lawson1957some}. This results in the necessary condition $\partial T/\partial t>0$ at the time of minimum hot-spot volume (stagnation time), where $T$ is the fuel temperature. However, this static analysis does not account for work done by the expansion of the fuel. Inclusion of this \cite{Springer_2018, patel2020hotspot} results in an extra condition $\partial^2 T/\partial t^2>0$ at stagnation time. The resulting dynamic ignition threshold is somewhat beyond the simpler static threshold \cite{Springer_2018}. 

Although almost spherically symmetric, inertial confinement fusion fuel capsules have perturbations that grow during the implosion \cite{clark2015radiation}. This is due to the short timescales for fluid instabilities. Power imbalance between different laser beams can also create these fuel asymmetries \cite{schlossberg2021observation}. Multi-dimensional ICF modeling has mostly focussed on yield degradation from the resulting change in hot-spot shape, both with low spatial wavenumber \cite{hurricane2020analytic} and high wavenumber \cite{haines2017high} asymmetries. Asymmetry also prevents efficient conversion of the implosion energy into hot-spot thermal energy \cite{casey2021evidence}. 
Ablative stabilization of the fluid instabilities typically leads to a dominant unstable mode during implosion, producing $5-10$ micron-scale spikes entering the hot-spot at stagnation time \cite{takabe1985self, clark2015radiation, haines2020cross}. Engineering features are often the largest target perturbations, producing jets that penetrate the stagnated fuel at close to the implosion velocity \cite{haines2020cross, pak2020impact}.

Although the yield reduction due to fuel asymmetry is well studied, the asymmetries can also lead to additional plasma physics considerations. In this work, we estimate the impact of magnetic fields $\mathbf B$ that are self-generated by the spikes. If the temperature and density gradients are misaligned then the Debye sheath electric field has a curl, producing long-lasting magnetic fields via Faraday's law \cite{biermann1950ursprung}. This Biermann magnetic field is then advected with the fluid motion, heat flux and currents \cite{braginskii1958transport}.

\begin{figure*}[t!]
  \centering
  \includegraphics[]{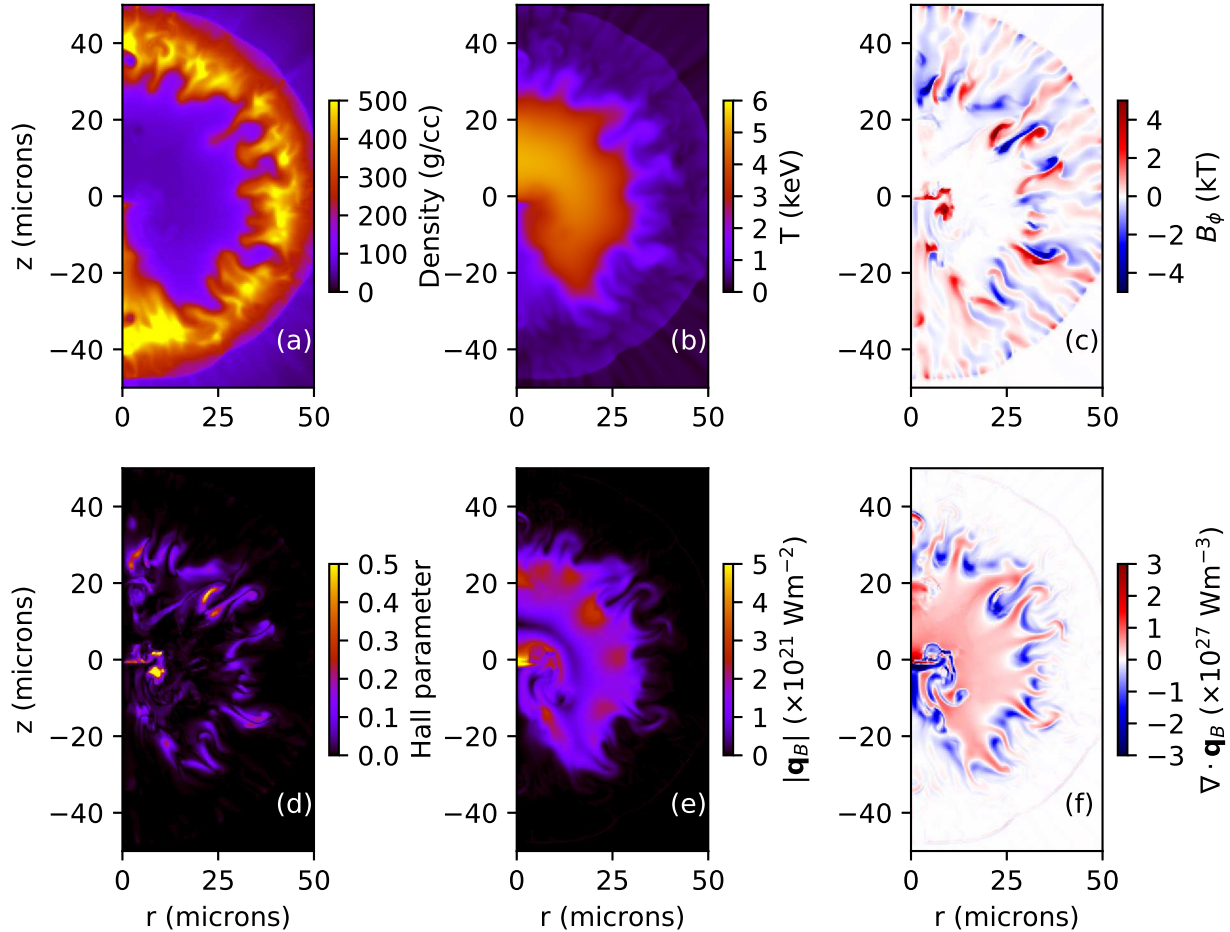}
  \caption{ Results from the xRAGE radiation-hydrodynamic simulation in two-dimensional cylindrical geometry, as well as XMHD post-processing to find the self-generated magnetic fields and the resulting reduction of heat loss. Results are shown near to stagnation time. Panels show the (a) mass density, (b) electron temperature, (c)  self-generated magnetic field (which is purely azimuthal in this cylindrical geometry), (d) the resulting Hall parameter, (e) the magnitude of the magnetized heat flux and (f) the divergence of the magnetized heat flux. }
  \label{xrage}
\end{figure*}

Numerical simulations have  predicted $|\mathbf B|\simeq 5\,$kT in current national ignition facility (NIF) experiments \cite{walsh2017self}. Separate post-processing \cite{sadler2020magnetization} of radiation-hydrodynamic simulations with the xRAGE code was in agreement with this value. Recently, a much simpler and faster post-processing technique was developed \cite{ walsh2021biermann}. It numerically integrates the Biermann magnetic flux from the time dependent fuel temperature, density and asymmetry. For cases with minimal asymmetry amplitude, it was found to be in good agreement with the previous more detailed methods. It can be applied, with some assumptions, across the whole ignition parameter space. In this work, we use this new self-magnetization model to estimate changes to hot-spot implosion trajectories and the ignition threshold.

Although far beyond typical laboratory field strengths, the $|\mathbf B|\simeq 5\,$kT value is still too small to directly affect hot-spot hydrodynamics through the magnetic pressure and tension \cite{walsh2017self}. However, magnetic fields could still indirectly influence the hydrodynamics and ignition threshold by insulating the electron heat flux, a key route of hot-spot energy loss. The $7\,$fs electron gyro-period at $|\mathbf B|=5\,$kT is similar to the electron Coulomb collision time in the hot-spot. As such, heat transport perpendicular to $\mathbf B$ is heavily reduced and deflected \cite{braginskii1958transport}. 

This effect has been measured to improve direct drive ICF yield (even well below the ignition threshold) by insulating the hot fuel with an external magnetic field \cite{chang2011fusion, hohenberger2012inertial}. The magnetized liner inertial fusion concept \cite{slutz2010pulsed} also uses external magnetization to this effect. With sufficient compression, the magnetic field can even confine fast alpha particles from the fusion reactions. The question remains - if external fields are known to improve fusion yield, to what extent are the self-generated fields improving the performance of standard ICF?

To answer this question, in section two we extend our previous analysis of the data from a two-dimensional radiation-hydrodynamic simulation \cite{sadler2020magnetization}. We now go further, and use the post-processed magnetic field profile to quantify the reduction in hot-spot heat loss. In section three, we develop a self-magnetization model for ICF, as a function of the hot-spot temperature and areal density. We use this to draw static ignition thresholds with and without the self-magnetization. In section four, we show how the self-magnetization model can be included into the hot-spot energy balance equations, allowing a dynamic self-magnetized hot-spot model. We show that self-magnetization effects should increase strongly with temperature, becoming a key consideration in the record burning plasmas of recent NIF experiments\cite{kritcher2022design, zylstra2022burning, N210808press}. Section five provides a conclusion and discussion of assumptions.

\section{Analysis of a two-dimensional simulation}
The raw xRAGE radiation-hydrodynamic simulation data is shown in Fig. \ref{xrage}. The simulation used a two-dimensional Eulerian cylindrical geometry with adaptive mesh refinement of minimal cell size $0.25\mu$m. The setup had realistic capsule and X-ray drive perturbations, emulating the measured\cite{le2018fusion} target parameters of NIF shot N170601. This used a $980\,\mu$m outer radius capsule with high density carbon ablator. Further simulation details are given in ref. \cite{haines2020cross}. The setup also modeled the fuel fill tube engineering feature, which tends to produce a jet of contaminants into the hot-spot. As Biermann magnetic fields primarily grow from these asymmetries, the use of detailed initial target metrology is a pre-requisite for accurately simulating the $\mathbf B$ profile. 

Figs. \ref{xrage}a-b show the plasma density and temperature at stagnation time. The dense shell region encloses the hot-spot, which is at much lower density and higher temperature. The fill tube jet enters along the $z$ axis, composed partly of carbon. The dense spikes mentioned in the introduction are entering around the edge of the hot-spot. The total mass is $210\,\mu$g.

By integrating the XMHD model \cite{sadler2021symmetric}, we can estimate the magnitude and profile of the Biermann magnetic fields from the radiation hydrodynamic data. This analysis was described in ref. \cite{sadler2020magnetization}, and the result is shown in Fig. \ref{xrage}c. Due to the axisymmetric two-dimensional simulation, the field direction is purely azimuthal. Resistance from Coulomb collisions results in diffusion of the magnetic fields, with a length scale $\simeq 1\,\mu$m. The model has been evolved in time using a two-dimensional flux corrected transport approach, on a uniform grid with resolution $0.25\,\mu$m. The numerical integration covered only the $300\,$ps period preceding stagnation, and so it may under-estimate the magnetic flux accumulated during the full implosion. 

Magnetic fields begin to significantly affect heat flux when the dimensionless electron Hall parameter 
\begin{align}
    \chi = \frac{e|\mathbf{B}|\tau_{ei}}{m_e} \simeq \frac{10^{-7}A}{ Z^2\ln(\Lambda)}\left(\frac{|\mathbf{B}|}{\mathrm{T}}\right)\left(\frac{T}{\mathrm{eV}}\right)^{3/2}\left(\frac{\rho}{\mathrm{gcm}^{-3}}\right)^{-1}\label{chi}
\end{align}
approaches 1, where $e$ is the electron charge, $\tau_{ei}$ is the Braginskii electron-ion Coulomb collision time, $m_e$ is the electron mass, $A$ is the ion mass number, $Z$ is the ion charge, $\ln(\Lambda)$ is the Coulomb logarithm, $T$ is the temperature and $\rho$ is the mass density. The resulting Hall parameter profile is shown in Fig. \ref{xrage}d. The maximal value exceeds $0.5$, although it is closer to $0.15$ across much of the hot-spot edge. 

\begin{figure}[t!]
  \centering
  \includegraphics[]{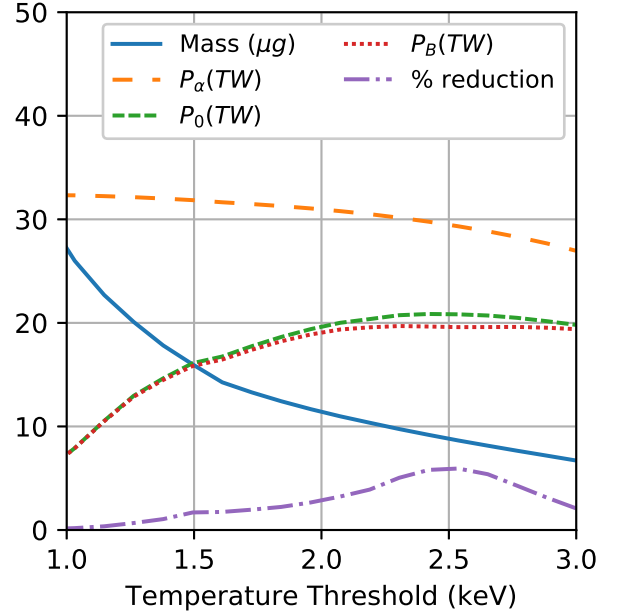}
  \caption{ Post-processed quantities as a function of the hot-spot temperature threshold definition, for the xRAGE data shown in Fig. \ref{xrage}. Lines show the enclosed hot-spot mass (solid), the fusion $\alpha$ power generated inside the enclosed volume (long dashes), the unmagnetized heat loss (short dashes), the magnetized heat loss (dotted), and the percentage reduction in the heat loss due to self-magnetization (dash-dot). }
  \label{threshold}
\end{figure}

The magnetized heat flux $|\mathbf{q}_B|$ is shown in Fig. \ref{xrage}e. It has been calculated using the transport coefficient theory of ref. \cite{epperlein1986plasma}. Its magnitude is reduced in regions with large $\chi$. With the divergence theorem, the integral calculation $P_{B}=\int \nabla\cdot\mathbf q_B\,d^3\mathbf x$ provides the hot-spot heat loss. The integral is taken over the hot-spot volume, defined as the region with $T>T_{th}$. The unmagnetized heat loss $P_0$ can be found using the same calculation with $\chi=0$. Fig. \ref{threshold} shows these as a function of the definition of the hot-spot threshold $T_{th}$. The percentage heat loss reduction due to self-magnetization $(P_0-P_B)/P_0$ is also shown in Fig. \ref{threshold}, peaking at around $6\%$. Both the heat loss and its reduction due to the magnetic field are maximal for $T_{th}\simeq 2.5\,$keV. The heat loss of $\simeq 20\,$TW is similar to the fusion $\alpha$ particle power $P_\alpha\simeq 30\,$TW. This $2.5\,$keV threshold can be compared with Figs. \ref{xrage}b-d, for which the maximal Hall parameter values are located around $T\simeq 2.5\,$keV at the hot-spot edge. The mean of the temperature weighted by Hall parameter is $(\int\,d^3\mathbf{x}\, T\chi)/(\int\,d^3\mathbf{x}\,\chi) = 2.3\,$keV, around 43\% of the peak temperature $5.3\,$keV and 60\% of the burn averaged temperature $3.9\,$keV. 

There are several uncertainties involved in using a post-processing method to evaluate magnetic reduction in heat loss. In reality, the reduced heat conductivity will steepen the temperature gradient. This means the self-consistent heat loss reduction is likely less than the $6\%$ value shown in Fig. \ref{threshold}. However, a steeper temperature gradient will also increase the Biermann magnetic field generation. The true reduction is therefore not simply related to the post-processing analysis shown here. An additional uncertainty is the role of magnetic fields in three dimensions, versus the two-dimensional simulation shown here.

A common model\cite{chang2011fusion} for magnetized hot-spot heat loss reduction is
\begin{align}
    \frac{P_0-P_B}{P_0} \simeq f_B\left(1-\frac{\kappa_\perp(\chi)}{\kappa_\perp(0)}\right),\label{reduction}
\end{align}
where $\kappa_\perp(\chi)$ is the magnetized heat conductivity coefficient \cite{epperlein1986plasma} for ionization $Z=1$. This model reflects the fact that Biermann magnetic fields do not cover the full surface of the hot-spot. It is clear from Fig. \ref{xrage}d that not all of the hot-spot edge is magnetized. The estimated surface filling factor is taken as $f_B=2/3$. This is motivated by the theory for the application of a linear external magnetic field \cite{walsh2021magnetized}, for which $f_B=2/3$. This is because $\mathbf B$ has a component through the hot-spot surface, and the heat flux parallel to $\mathbf B$ is unaffected. With use of $\chi=0.15$ (the hot-spot edge value in Fig. \ref{xrage}d), this model produces $(P_0-P_B)/P_0=5.7\,\%$, in agreement with the peak volume integrated reduction in Fig. \ref{threshold}. The value of $f_B=2/3$ therefore seems appropriate to match the detailed post-processing analysis, although this could change in three dimensions.

The magnetic field also deflects the heat flux, creating the Righi-Leduc component $\mathbf{q}_{RL}$. This has no effect on the heat loss in Eq. (\ref{reduction}), since $\mathbf{q}_{RL}\propto \nabla T\times \mathbf B$ is parallel to the hot-spot surface isotherm. This is why the Righi-Leduc transport coefficient $\kappa_\wedge$ is not included. However, with a self-consistent XMHD treatment\cite{walsh2021updated}, $\mathbf{q}_{RL}$ can funnel heat towards the spike bases and change their growth. This could have an indirect effect on the resulting Biermann generation and heat losses. 

Eq. (\ref{reduction}) still depends on $\chi$. In the next section, we develop a model to estimate the Hall parameter $\chi$ at stagnation, and thus close the self-magnetized ignition model. This estimate is found from first principles and it does not require multi-dimensional simulation data. It produces a result that agrees with the post-processed Hall parameter in Fig. \ref{xrage}d. Intriguingly, the model depends on the standard Lawson parameters - the stagnated hot-spot temperature and areal density. We emphasize that the $\simeq 6\%$ heat loss reduction in Fig. \ref{threshold} is evaluated using a shot that had a relatively low yield of $\simeq 50\,$kJ. The model suggests that the heat loss reduction should strongly increase in higher yield implosions.

\section{Magnetization model and the static ignition threshold}

To close the model for a general implosion design, the heat loss reduction factor in Eq. (\ref{reduction}) still needs an estimated Hall parameter $\chi$, which will be time-dependent. However, this can be estimated from the recent self-magnetization model of Walsh and Clark \cite{walsh2021biermann}

\begin{align}
    \frac{d\Phi}{dt} = \ln\left(\frac{\rho_{s}}{\rho_{hs}}\right)
\frac{T_{hs}-T_{s}}{e}\left|\frac{\Delta(\rho R) }{\langle\rho R\rangle} - \frac{\Delta(T R)}{\langle TR\rangle }\right|\tilde k.\label{walshmodel}
\end{align}
In this expression, $\Phi(t)$ is the magnetic flux, $T_{hs}(t)$ is the hot-spot temperature, $\rho_{hs}(t)$ is the hot-spot density, $T_s(t)$ is the shell temperature, and $\rho_s(t)$ is the shell density. Hot-spot quantities are calculated with a burn-weighted volume average. For the data in Fig. \ref{xrage}, this results in $T_{hs}=3.9\,$keV and $\rho_{hs}=90\,$gcm$^{-3}$. Working now in spherical coordinates, the quantities $\rho R=\int\,\rho(t, r, \theta)\,dr$ and $TR=\int\,T(t, r, \theta)\,dr$ are the density and temperature radially integrated over the hot-spot region only. Angle brackets represent an average over the polar angle $\theta$ with respect to the $z$ axis, given by $\langle \rho R\rangle=(1/\pi)\int_0^\pi \rho R\,d\theta$. Eq. (\ref{walshmodel}) represents the flux growth from a single perturbation mode number $\tilde k=2\pi R/\lambda$, where $R=\langle\rho R\rangle/\rho_{hs}$ is the hot-spot radius and $\lambda$ is the perturbation wavelength. Perturbation amplitudes are defined as the complex Fourier amplitudes $\Delta(\rho R) =  (1/\pi) \int_{0}^{2\pi} (\rho R)\exp{(i\tilde{k}\theta)}\,d\theta$ and $\Delta(T R) =  (1/\pi) \int_{0}^{2\pi} (T R)\exp{(i\tilde{k}\theta)}\,d\theta$ for the mode number $\tilde k$. Since the simulation has azimuthal symmetry, $(\rho R)(2\pi-\theta)=(\rho R)(\theta)$, the amplitudes are real in this case. So long as the perturbation amplitude remains small, previous work \cite{walsh2021biermann} found good agreement between multi-dimensional XMHD simulations and this post-processing model.

The definition of flux $\Phi$ needs careful consideration. The standard definition is $\Phi_0=\int_S \mathbf{B}\cdot d^2\mathbf x$, where $S$ is an open surface. However, inserting Faraday's law yields $\partial\Phi_0/\partial t = -\int (\nabla\times\mathbf E)\cdot d^2\mathbf x = -\int_C\mathbf{E}\cdot d\mathbf{l}$, where $d\mathbf{l}$ is the line element of the contour $C$ bounding $S$ and $\mathbf{E}$ is the electric field from the generalized Ohm's law. If $S$ is taken as the half-slice plane through the plasma with $r\geq 0$ (i.e. the full simulation plane shown in Fig. \ref{xrage}), the plasma has $\mathbf E =0$ on the three sides of the rectangle at far distances. Therefore $\Phi_0$ only contains information about the $z$ axis, and it has no information about plasma perturbations at other locations. For this reason, Walsh and Clark instead defined the flux as $\Phi=\int_{-\infty}^\infty\int_0^\infty |\mathbf{B}|\,dzdr$ in cylindrical coordinates (i.e. the scalar integral of the modulus of $\mathbf B$ over this same surface $S$).

\begin{figure}[t!]
  \centering
  \includegraphics[]{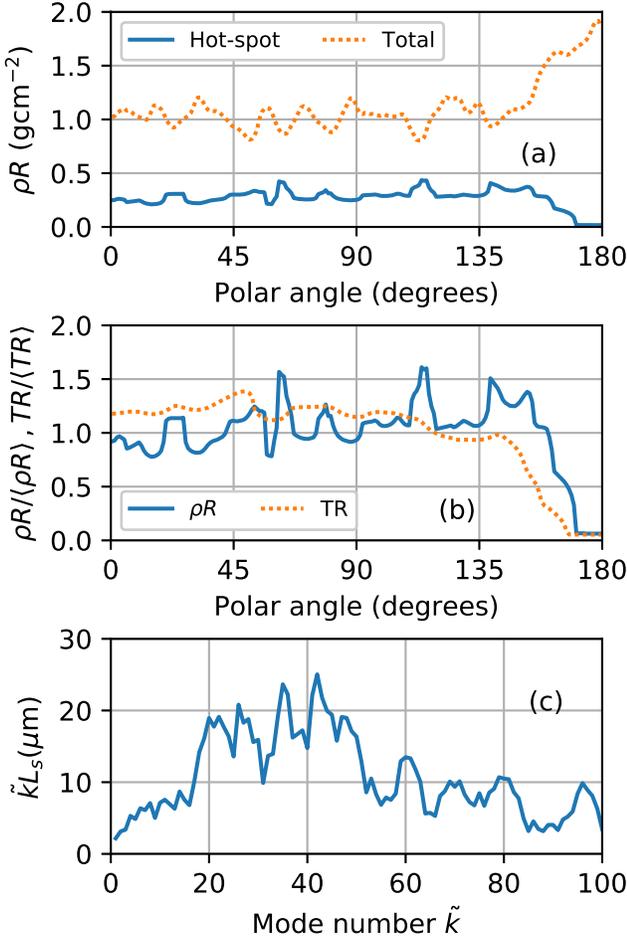}
  \caption{ Angular analysis of the xRAGE simulation data at the time-step shown in Fig. \ref{xrage}. (a) Areal density $\int\rho(r,\theta)\,dr$ as a function of polar angle $\theta$ from the $z$ axis. The dotted line shows the total areal density, and the solid line shows the numerical integral covering only the region inside the shell with $\rho<180\,$gcm$^{-3}$. The fill tube jet is visible near $\theta\simeq 180^\circ$. (b) Density and temperature radially integrated across the hot-spot region, normalized by their mean values. (c) Spectrum of the angular part of the magnetic source term given in Eqs. (\ref{walshmodel}-\ref{Lsdefinition}), as a function of mode number $\tilde k$. The resulting Biermann magnetic fields have a spectral peak at around $\tilde k=20-40$. The spectrum has been smoothed with a uniform kernal of width 5.  }
  \label{rhor}
\end{figure}

The variation of $\rho R$ with polar angle is shown in Fig. \ref{rhor}a for the xRAGE simulation. Two curves are shown - one for the numerical integral covering the full simulation domain, and one where it only covers the region inside the shell with $\rho<180\,$gcm$^{-3}$. The large fill tube jet is visible in the data at $\theta\simeq 180^\circ$, and several prominent spikes in Fig. \ref{xrage}a are also visible. In the spike positions, the total areal density is greater, but the hot-spot areal density is less. The average hot-spot areal density is $\langle\rho R\rangle=0.27\,$gcm$^{-2}$, giving $R=30\,\mu$m. Fig. \ref{rhor}b shows the radially integrated hot-spot density and temperature, normalized by their mean values. The temperature integral is taken over the region with $T>2.5\,$keV. Some correlation is visible between the density and temperature curves. Both have a lower value along the directions where spikes enter the hot-spot. However, due to thermal conduction, the temperature perturbations are much smoother. The hot-spot can be defined either by a density contour or a temperature contour. The differences in the shape of these contours leads to $\nabla T\times\nabla\rho\neq 0$ and Biermann magnetic field generation.

In Eq. (\ref{walshmodel}), the angular contribution to the magnetic flux growth is proportional to $\tilde k L_s$, where we define the quantity
\begin{align}
    L_s=R\left|\frac{\Delta(\rho R) }{\langle\rho R\rangle } - \frac{\Delta(T R) }{\langle TR\rangle }\right|,\label{Lsdefinition}
\end{align}
which is related to the length of the spikes and the differences between the temperature and density contours.
The angular contribution $\tilde k L_s$ for the xRAGE data is shown in Fig. \ref{rhor}c, for a variety of mode numbers.
 For a multi-mode perturbation, the flux calculation must retain this spectral information\cite{walsh2021biermann}. Due to ablative stabilization, the Biermann fields have a dominant spectral mode number \cite{walsh2021biermann}. There is a clear spectral peak in Fig. \ref{rhor}c for $\tilde k\simeq 20-40$. Smaller perturbations are suppressed by ablation. In addition, small scale Biermann magnetic fields tend to be dissipated by resistive diffusion. The multi-mode flux, at least for typical perturbation spectra in NIF implosions, can therefore be estimated by replacing the multi-mode spectrum with a single mode of $\tilde k\simeq 20$.  Of course, this perturbation spectral peak will depend on ablation physics amongst other things, and so $\tilde k$ is another free parameter that has significant uncertainty. Walsh and Clark\cite{walsh2021biermann} also found a spectral peak $\tilde k\simeq 20$ in XMHD simulations.
 
 The amplitude of the equivalent single mode can be approximated by the perturbation amplitude in Fig. \ref{rhor}b. Taking this data, and excluding the large variation in the fill tube jet, the relative perturbation amplitudes are $\sqrt{\langle(\rho R)^2\rangle - \langle\rho R\rangle^2}/\langle\rho R\rangle\simeq16\%$ and $\sqrt{\langle(T R)^2\rangle - \langle T R\rangle^2}/\langle T R\rangle\simeq 9\%$. Subtracting these, this gives an estimated $L_s\simeq 0.07\times30\,\mu$m $=2.1\,\mu$m.

Since Eq. (\ref{walshmodel}) is proportional to the temperature difference between the hot-spot and the shell, magnetic flux accumulates predominantly around stagnation time. This was confirmed with analysis of $\Phi(t)$ curves from XMHD simulations \cite{walsh2021biermann}. An estimate of the flux at stagnation is therefore given by $\Phi = \tau_{stag} d\Phi/dt$, where $d\Phi/dt$ is evaluated at stagnation and $\tau_{stag}$ the stagnation duration. This can be approximated \cite{hurricane2020analytic} by $\tau_{stag}=\sqrt{R/\ddot R}$, where the shell acceleration is given by Newton's law $m_s\ddot R=4\pi R^2p_{stag}$. Here, $m_s$ is the shell mass and $p_{stag}$ is the stagnation pressure. The magnetic flux at stagnation is then approximately
\begin{align}
    \Phi \simeq \frac{d\Phi}{dt}\tau_{stag} = 
     \tilde kL_s\ln\left(\frac{\rho_s}{\rho_{hs}}\right)\frac{T_{hs}}{e}\sqrt{\frac{m_s}{4\pi R^3 p_{stag}}},
\end{align}
where we have assumed that $T_{hs}\gg T_s$ near stagnation. Combining this relation with the ideal gas law, and using the hot-spot mass $m_{hs}=4\pi R^3\rho_{hs}/3$, leads to
\begin{align}
    \Phi \simeq 65\tilde k \sqrt{\frac{m_s}{m_{hs}}}\ln\left(\frac{\rho_s}{\rho_{hs}}\right)\left(\frac{L_s}{\mathrm{m}}\right)\left(\frac{T_{hs}}{\mathrm{eV}}\right)^{\frac{1}{2}}\,\mu\mathrm{Tm}^2.\label{phimodel}
\end{align}
The accumulated flux increases with the shell to hot-spot mass ratio and with the hot-spot temperature. As expected, it also increases with the mode number and amplitude of the asymmetries. With use of $T_{hs}= 3.9\,$keV, $\tilde k=20$, $L_s\simeq 2.1\,\mu$m, and the masses and densities from Figs. \ref{xrage}-\ref{threshold} ($\rho_s=400\,$gcm$^{-3}$, $\rho_{hs}=90\,$gcm$^{-3}$, $m_{hs}=10\,\mu$g, $m_{s}=200\,\mu$g),  Eq. (\ref{phimodel}) leads to $\Phi \simeq 1.1\,\mu$Tm$^2$. This compares with the numerically integrated value of $\Phi=1.56\,\mu$Tm$^2$ for the post-processed field profile in Fig. \ref{xrage}c. 

To close the model and use $\Phi$ to find an approximate value for $\chi$, we must also make an assumption for the area containing the magnetic field. Referring to Fig. \ref{xrage}c, the area containing significant field strength is far less than the hot-spot area. We take this as a semi-circle annulus with area $\pi R\Delta r_B$, where $R$ is the hot-spot radius and $\Delta r_B$ is the width of the magnetic field region at the edge of the hot-spot. Due to Nernst advection of the magnetic fields to the hot-spot edge, this width is far less than $R$. Due to the definition of $\Phi$ as the area integral of $|\mathbf B|$, the field strength can then be simply estimated as $|\mathbf{B}|\simeq \Phi/(\pi R\Delta r_B)$.

Finally, the model can be closed by finding $\chi$ from Eq. (\ref{chi}). Since the magnetic field is primarily generated around the hot-spot edge, and is kept there by Nernst advection, we use the hot-spot edge temperature $T_{th}$ for this calculation, rather than the burn-averaged hot-spot temperature. In the xRAGE data, this was $T_{th}\simeq 0.65T_{hs}$, where $T_{hs}$ is the burn averaged temperature. Similarly, the density at this interface position is estimated as $\rho_{hs}/0.65$. Taking a fixed value $\ln(\Lambda)=5$, these values result in the estimated Hall parameter for self-generated magnetic fields
\begin{align}
    \chi\simeq c_\chi \tilde k \ln\left(\frac{\rho_s}{\rho_{hs}}\right)\sqrt{\frac{m_s}{m_{hs}}}\frac{L_s}{\Delta r_B}\left(\frac{T_{hs}}{\mathrm{keV}}\right)^2\left(\frac{\langle\rho R\rangle_{hs}}{\mathrm{gcm}^{-2}}\right)^{-1},\label{chimodel}
\end{align}
where $c_\chi\simeq 2\times 10^{-5}$ and all quantities are evaluated at bang time (time of maximum fusion rate). Due to its dependence on $T_{hs}$ and $\langle\rho R\rangle_{hs}$, this form is appropriate for inclusion into a Lawson-type analysis. Eq. (\ref{chimodel}) shows that, in keeping with prior expectations, magnetization effects should increase with temperature. Furthermore, they decrease with hot-spot areal density. This is because denser plasma is harder to magnetize [Eq. (\ref{chi})], and a greater stagnation radius means less compression of the accumulated magnetic flux. The uncertainty in the parameters $L_s$ and $\Delta r_B$ is the main source of error in this simple model. However, the model only depends on their ratio. With reference to Fig. \ref{xrage}a-d, a physically reasonable estimate for prominent spikes is to take $\Delta r_B = L_s$. It should be noted that in the limit of small perturbations, the magnetic field profile is instead set by resistive diffusion, and so $L_s\ll\Delta r_B$. In this limit, Eq. (\ref{chimodel}) therefore tends to zero, remaining consistent with the physical intuition. Evaluation of Eq. (\ref{chimodel}) with $\Delta r_B\simeq L_s$ and the values from Fig. \ref{xrage}a-b ($\tilde k=20$, $m_{hs}=10\,\mu$g, $m_{s}=200\,\mu$g, $\rho_{s}=400\,$gcm$^{-3}$, $\rho_{hs}=90\,$gcm$^{-3}$, $T_{hs}=3.9\,$keV, $\langle\rho R\rangle_{hs}=0.27\,$gcm$^{-2}$) results in $\chi\simeq 0.15$. This is similar to the values around the hot-spot edge in Fig. \ref{xrage}d.

\begin{figure}[t!]
  \centering
  \includegraphics[]{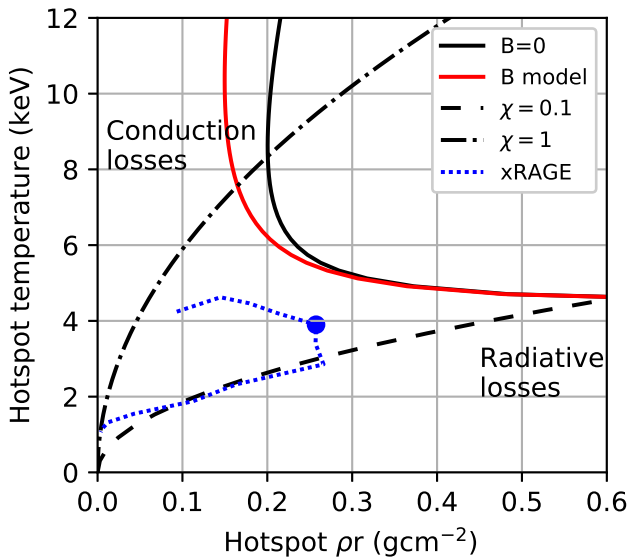}
  \caption{ Static Lawson ignition criteria for the case with no self-magnetization (solid black), and with the self-magnetization model [Eq. \ref{chimodel}] (solid red). Contours of this model are shown for $\chi=0.1$ (dashed) and $\chi=1$ (dash-dot). The simulated (clockwise) trajectory of shot N170601 is shown in dotted blue, calculated from the xRAGE simulation shown in Fig. \ref{xrage}. These are shown as a function of the hot-spot areal density and the burn averaged hot-spot temperature. }
  \label{lawsonchi}
\end{figure}

Contours of Eq. (\ref{chimodel}) are shown in the Lawson plane in Fig. \ref{lawsonchi}, for the typical case $\rho_s/\rho_{hs}=5$, $m_s/m_{hs}=20$, dominant mode number $\tilde k=20$, and $\Delta r_B=L_s$. We again note that this is the expected value for $\chi$ across large regions of the hot-spot edge. As in Fig. \ref{xrage}d, there may be localized regions with even greater $\chi$ values, perhaps by a factor of $5-10$. The trajectory for the xRAGE simulation is shown in dotted blue, with the snapshot in Fig. \ref{xrage} shown by the dot.

There is significant uncertainty in this self magnetization model, predominantly 
originating from the magnetic field profile effects encapsulated here in $\Delta r_B$, $L_s$ and $f_B$. It appears that $L_s\simeq\Delta r_B$ is a good estimate, at least for the comparison with the simulation data in Fig. \ref{xrage}. Multi-dimensional effects or the self-consistent feedback of the magnetized insulation into the hydrodynamic development could invalidate the model. As such, an accurate validation of the Hall parameter model and ignition threshold will require a large scale parameter scan with multi-dimensional XMHD simulations. This is left to future work. For now, we assume fixed values $f_B=2/3$ and $\Delta r_B=L_s$. Additional uncertainty arises from the validity of the XMHD and Biermann models themselves. For steep temperature gradients, Biermann generation is heavily reduced\cite{ridgers2021inadequacy, sherlock2020suppression}. This kinetic reduction could be important in ICF, especially earlier on in the implosion and at the higher temperatures typical of ignition.

Keeping these limitations in mind, a simple ignition threshold can now be obtained by considering the hot-spot energy balance \cite{lawson1957some}. In addition to the magnetized heat conduction, radiation also causes significant energy loss. Assuming 50/50 Deuterium-Tritium fuel, the classical continuum Bremsstrahlung power is $P_{rad}= C_Bn_e^2\sqrt{T_{hs}}Ve^{-\sigma}$, where $n_e=\rho_{hs}/m_i$ is the electron number density, $m_i=4.2\times10^{-27}\,$kg is the average deuterium-tritium ion mass, $V$ is the hot-spot volume and $C_B = 1.69\times 10^{-38}\,$Wm$^3$eV$^{-1/2}$. Without the $\sigma$ optical depth term included, the $dT_{hs}/dt\propto -\sqrt{T_{hs}}$ dependence will cause temperature to fall to zero after some finite time. To prevent this unphysical behavior, we use the model \cite{befki1966radiation} $\sigma= C_dn_e^2RT_{hs}^{-7/2}$ with $C_d=6\times10^{-48}$m$^5eV^{7/2}$.

To achieve ignition, a necessary criterion is that conduction and radiative losses must be exceeded by the power of fusion alpha particles $P_\mathrm\alpha = 0.25n_e^2 F(T_{hs})VE_\alpha$. In this expression, $F(T_{hs})$ is the DT fusion reactivity and $E_\alpha=3.5\,$MeV is the fusion alpha particle energy. The alpha particle deposition fraction $\theta_\alpha(\langle\rho R\rangle_{hs})$ gives the fraction of this alpha particle energy that is deposited in the hot-spot \cite{krokhin1973escape, christopherson2020theory}. Finally, taking the temperature scale-length as $R$ and fixed $\ln(\Lambda)=5$, the Spitzer model yields $P_0 = C_c T_{hs}^{7/2}R$, with $C_c = 77000\,$WeV$^{-7/2}$m$^{-1}$.

The numerical solution to $P_\mathrm\alpha \theta_\alpha=P_0+P_\mathrm{rad}$ is the black line in Fig. \ref{lawsonchi}, in agreement with the curve in ref. \cite{hurricane2018beyond}. Radiative losses create a minimum ignition temperature $T_{hs}>T_\mathrm{ideal}\simeq 4.5\,$keV, unless the fuel is so large that it becomes optically thick. For lower areal density values, conduction losses become important too, leading to even higher required temperatures. 

As discussed and justified in the previous section, we take the magnetized heat loss model $P_B = P_0[1+2\kappa_\perp(\chi)/\kappa_\perp(0)]/3$. The $\chi$ value is taken from the model in Eq. (\ref{chimodel}). The numerical solution to $P_\mathrm\alpha \theta_\alpha=P_B+P_\mathrm{rad}$ is shown solid red in Fig. \ref{lawsonchi}. The radiative loss, and therefore the ignition requirement $T_{hs}>T_\mathrm{ideal}$, are unaffected. For NIF stagnation parameters $T_{hs}\simeq 5.5\,$keV and $\langle\rho R\rangle_{hs}\simeq 0.25\,$gcm$^{-2}$, the model gives an ignition threshold reduction $\simeq 200\,$eV. The magnetization effect becomes even more pronounced at higher temperatures. The additional heat confinement reduces the required hot-spot areal density from $0.2\,$gcm$^{-2}$ to $0.15\,$gcm$^{-2}$.

In agreement with the quenching in the xRAGE simulation, the stagnated hot-spot temperature is below both of the static ignition curves. However, the stagnation is notably closer to the magnetized curve than it is to the unmagnetized curve. 

\section{Dynamic energy balance model}

The static ignition threshold has the limitation that it does not consider expansion work done. To demonstrate the time-dependent effects of self-magnetization, we now numerically integrate a simple hot-spot energy balance model. In a time dependent model with variable hot-spot radius $R$, hot-spot energy $E$ will also change with work done. This results in 
\begin{align}
    \frac{dE}{dt} = P_\mathrm\alpha \theta_\alpha - P_B - P_\mathrm{rad} - p\frac{dV}{dt},\label{Ehs}
\end{align}
where $p=2E/(3V)$ is the total hot-spot pressure and $V=4\pi R^3/3$ is the hot-spot volume. This simple model assumes a spherical hot-spot. Shape perturbations have been discussed elsewhere \cite{hurricane2020analytic}.

We assume the simple thin-shell model, neglecting the fuel shell dynamics in the more complex model of ref. \cite{christopherson2020theory}. The total fuel mass $m_{tot}=m_{hs}+m_s$ is assumed constant. The hot-spot mass increases with ablation of the dense shell, where the energy source is conduction and the alpha particles that exit the hot-spot. The mass ablation is taken as
\begin{equation}
    \frac{dm_{hs}}{dt} = \frac{\rho_{hs}[(1-\theta_\alpha)P_\alpha + P_B]}{5p}.
\end{equation}

The hot-spot density is then found from $\rho_{hs}=m_{hs}/V$, whereas the thin shell density is $\rho_s=m_s/(4\pi R^2\Delta r_{s})$ with fixed shell thickness $\Delta r_s=20\,\mu$m. The development of the hot-spot size is given by Newton's law for the hot-spot pressure on the dense shell $m_s\ddot{R} = 4\pi R^2 p$.

\begin{figure}[t!]
  \centering
  \includegraphics[]{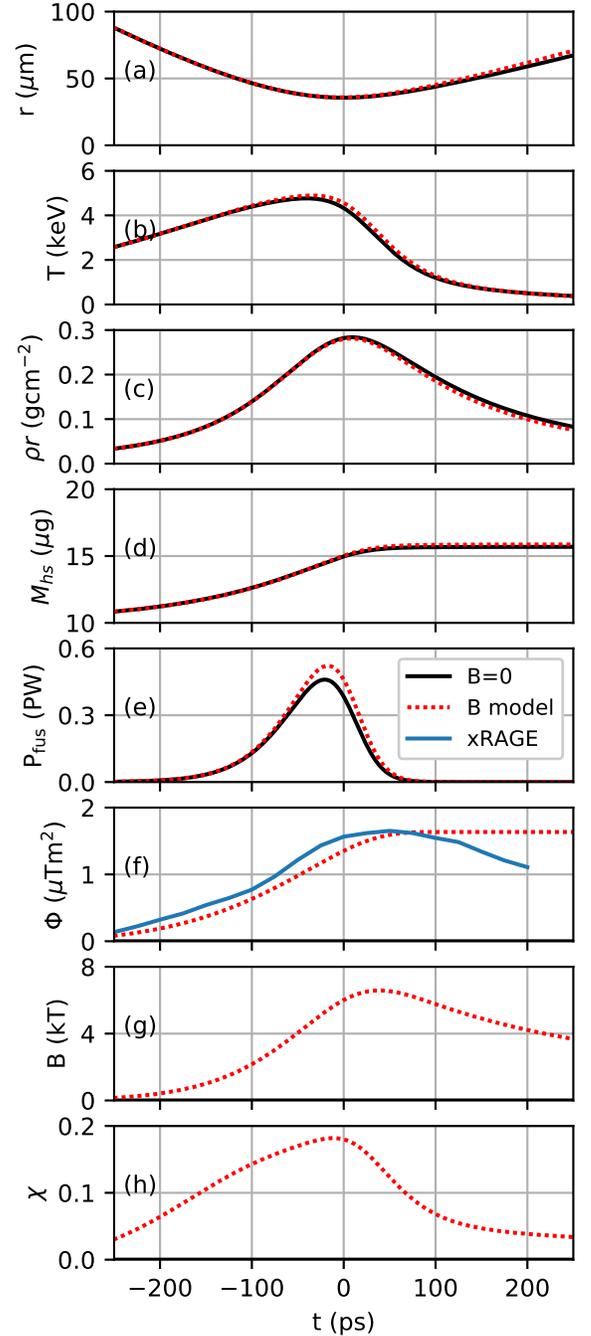}
  \caption{ Results of the time dependent hot-spot energy balance model for initial conditions $m_{tot}=200\,\mu$g, $m_{hs}=10\,\mu$g, $R=500\,\mu$m, $\dot R=-380\,$kms$^{-1}$ and $E=144\,$J. Numerical integration was repeated for the cases with the self-magnetization model (dotted red) and without (solid black). Panels show the time evolution of the hot-spot (a) radius, (b) temperature, (c) areal density, (d) mass, (e) fusion power, (f) magnetic flux, (g) magnetic field strength, and (h) Hall parameter. Time is given relative to stagnation of the unmagnetized case. The maximal heat loss reduction is $7.6\,\%$. The solid blue line in (f) shows the evolution of $\Phi$ in the xRAGE post-processing, with the frame shown in Fig. \ref{xrage} at $t=0$.}
  \label{model1}
\end{figure}

The model for self-magnetized heat loss $P_B$ was described in Eq. (\ref{reduction}). The magnetic flux is numerically integrated using Eq. (\ref{walshmodel}). A more advanced model should treat the perturbation amplitude as time dependent. Biermann growth from time-dependent perturbations was explored in ref. \cite{walsh2021biermann}. For now, we simply assume fixed values $\tilde k=20$, and $L_s=2\,\mu$m. Finally, $\chi$ is calculated at each time-step using $|\mathbf{B}|=\Phi/(\pi R L_s)$ and Eq. (\ref{chi}). Due to the field profile residing at the hot-spot edge, we use the values $0.4T_{hs}$ and $\rho_{hs}/0.4$ in the calculation of $\chi$ with Eq. (\ref{chi}). The shell temperature is assumed fixed at $T_s=1.5\,$keV, and $\Phi=0$ until $T_{hs}>T_s$.

Although this simple model lacks more detailed physics, it is sufficient to estimate the change in implosion trajectory due to self-magnetization. More advanced models should include effects such as shell dynamics and compressibility \cite{christopherson2018comprehensive}, radiation transport, fuel burn-up, spike dynamics, burn propagation, asymmetries \cite{hurricane2020analytic}, and kinetic effects. As such, a more accurate assessment of the ignition threshold will require a parameter scan with multi-dimensional perturbed radiation-XMHD simulations, something beyond the scope of this present initial work.

The model was numerically integrated over time using a second order Runge-Kutta method with time-step $0.5\,$ps. Fig. \ref{model1} shows the time dependent quantities. Initial conditions were chosen similar to those of the N170601 xRAGE simulation shown in Fig. \ref{xrage}. The resulting stagnation  radius $R=36\,\mu$m and areal density $0.28\,$gcm$^{-2}$ are similar to Fig. \ref{xrage} and Fig. \ref{rhor}. The temperature peaks slightly before stagnation with $T_{hs}\simeq 4.8\,$keV. The hot-spot mass increases from $10\,\mu$g to $16\,\mu$g. After stagnation, radiative losses cause a steep drop in temperature. Parameters were adjusted to match the fusion yield to the $47\,$kJ value from the xRAGE simulation \cite{haines2020cross} and the experiment\cite{le2018fusion}.

\begin{figure}[t!]
  \centering
  \includegraphics[]{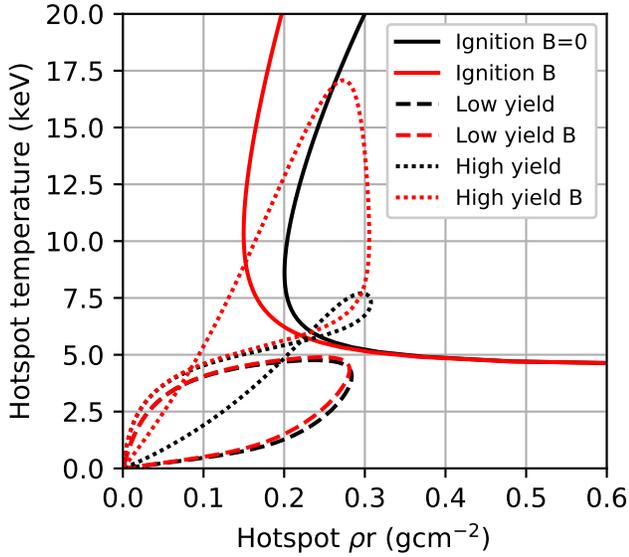}
  \caption{ Results of the time dependent hot-spot energy balance models for the cases shown in Fig. \ref{model1} (dashed, clockwise) and Fig. \ref{model2} (dotted, counter-clockwise). The solid lines show the static ignition thresholds reproduced from Fig. \ref{lawsonchi}. Red lines show the case with self-magnetization and black lines show the case without. }
  \label{lawson2}
\end{figure}

The numerical integration was repeated with the self-magnetization model. This is shown dotted red in Fig. \ref{model1}. The corresponding magnetic flux, magnetic field strength and Hall parameter at $t=0$ all agree with Fig. \ref{xrage}. The flux evolution is well matched by the time-dependent flux in the post-processing of the xRAGE simulation, perhaps with a slight time offset. The model does not predict the decrease at late times, which occurs due to resistive dissipation. Predominantly due to compression of the field profile, magnetic effects are strongest around the stagnation time. The Hall parameter is strongly linked to the hot-spot temperature evolution. It reaches a value of $0.18$, producing a heat loss reduction of 7.6\%. Use of Eq. (\ref{chimodel}) with the values from Fig. \ref{model1}a-d at bang time ($T_{hs}=4.8\,$keV, $\rho_{hs} R=0.26\,$gcm$^{-3}$, $m_{hs}=14\,\mu$g) results in $\chi=0.26$, in approximate agreement with the peak value in Fig. \ref{model1}h. The self-magnetization causes some subtle hydrodynamic differences, especially in the temperature and fusion power profiles. Yield is increased to $53.5\,$kJ. 

\begin{figure}[t!]
  \centering
  \includegraphics[]{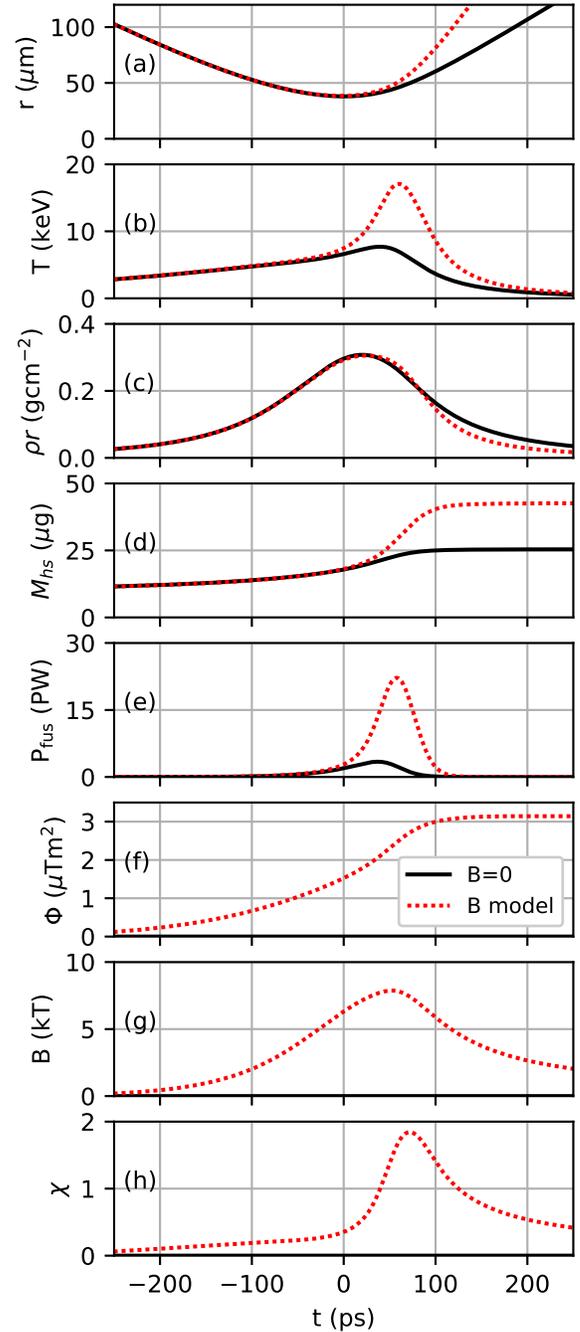}
  \caption{ Results of the time dependent hot-spot energy balance model for initial conditions $m_{tot}=200\,\mu$g, $m_{hs}=10\,\mu$g, $R=500\,\mu$m, $\dot R=-440\,$kms$^{-1}$ and $E=272\,$J. Numerical integration was repeated for the cases with the self-magnetization model (dotted red) and without (solid black). Panels show the time evolution of the hot-spot (a) radius, (b) temperature, (c) areal density, (d) mass, (e) fusion power, (f) magnetic flux, (g) magnetic field strength, and (h) Hall parameter. Time is given relative to stagnation of the unmagnetized case. The maximal heat loss reduction is $57\,\%$. }
  \label{model2}
\end{figure}

The model trajectories are also shown dashed in Fig. \ref{lawson2}. The curves are qualitatively similar to the xRAGE simulation curve in Fig. \ref{lawsonchi}. Differences are likely due to the heavy effects of asymmetries and the fill tube jet in the simulation, which cause a decrease in temperature of the hot-spot. Asymmetries are not included in the spherical zero-dimensional model. 

The effects of self-magnetization become more dramatic with proximity to ignition. For example, Fig. \ref{model2} shows the results of two runs with much higher yield. Again, the self-magnetized model is shown in red and the case with $\chi=0$ is shown in black. The initial conditions have a much higher implosion velocity. As such, the unmagnetized case reaches a maximal temperature of $7.7\,$keV. The areal density $\rho_{hs} R=0.31\,$gcm$^{-2}$ is similar to the previous case in Fig. \ref{model1}. The larger temperature causes a greater amount of ablation into the hot-spot, increasing the hot-spot mass to $25\,\mu$g. Burn now occurs predominantly after stagnation. The yield is $283\,$kJ. 

Due to the higher temperatures, magnetic heat insulation has a much greater effect now. There is a significant performance increase due to the heat insulation, and the self-magnetized yield is $1.3\,$MJ. The maximal temperature is $17\,$keV, and the burn averaged temperature is $13.5\,$keV. Although the yield is significantly higher than in Fig. \ref{model1}, the accumulated magnetic flux is only around double. Similarly, the increased stagnation radius means magnetic compression is not as effective, and the peak magnetic field is only slightly higher at $7.9\,$kT. However, magnetic insulation still increases dramatically because of the strong dependence of the Hall parameter on $T$ [Eq. (\ref{chi})]. The Hall parameter reaches values close to $2$, producing a heat loss reduction of 57\%. Some localized regions may have $\chi\simeq 5$ or more. Another key conclusion from Fig. \ref{model2} is that the maximal $\chi$ value occurs around bang time (time of peak fusion rate), which is after stagnation time (time of minimum volume) in this case. This is linked to the strong dependence of $\chi$ on the plasma temperature. 

This suggests that use of bang-time quantities in Eq. (\ref{chimodel}) should give the peak magnetization value. Using the values from Fig. \ref{model2}a-d at bang time ($\tilde k=20$, $T_{hs}=17\,$keV, $\rho_{hs} R=0.26\,$gcm$^{-2}$, $m_{hs}=30\,\mu$g, $\rho_s=270\,$gcm$^{-3}$, $\rho_{hs}=60\,$gcm$^{-3}$) results in $\chi\simeq 1.6$. This estimate is in agreement with the peak value $\chi\simeq 1.8$ in Fig. \ref{model2}h. Furthermore, the magnetization is dependent on the actual conditions in the hot-spot, rather than the hypothetical \emph{no alpha} quantities\cite{zhou2008measurable} that neglect fusion deposition, and are sometimes used in ignition threshold analysis.

Although the self-magnetization considerably increases with temperature, by this point in ignition the hot-spot energy balance and transport is dominated by alpha particles rather than electron heat conduction. However, if the magnetic field is sufficiently large, even the alpha particle transport could be confined. This will become important when $\Delta r_B/r_L>1$, where $r_L=270(|\mathbf B|/\mathrm{kT})^{-1}\,\mu$m is the Larmor radius of fusion alpha particles. This ratio is only around $0.1$ for the magnetized case in Fig. \ref{model2}, suggesting that alpha particles will not be magnetized until well into the propagating ignition regime. Detailed physics such as kinetic magnetized alpha particle and radiation transport \cite{appelbe2021magnetic} could significantly alter igniting plasma yields. 

\section{Summary}

We have used post-processing of fluid data to justify a model for the reduction of fusion fuel heat loss due to self-generated magnetic fields. For the post-processed simulation of HDC experiment N170601, where the spikes reached a length of $\simeq 5\,\mu$m, magnetic fields of $\simeq 5\,$kT surround the hot-spot. Integrating across the hot-spot boundary, heat loss is reduced by approximately 6\%. With some reasonable choices for its free parameters, the model matches this value. The predicted Hall parameter and heat loss reduction increase with temperature and decrease with areal density. The self-magnetization model reduced the ignition threshold temperature by $\simeq 200\,$eV for NIF experimental parameters.

There are several caveats to the conclusions reached in this work. The key additional effect is multi-dimensional in nature, with the interaction of the Biermann heat flux insulation with three-dimensional fluid instabilities. Isolated studies suggested that magnetized heat flux increases the spike growth rate \cite{matsuo2021enhancement, walsh2021updated}. This will likely reduce the fusion yield, especially if spikes are composed of highly radiative contaminants. A further consideration is the applicability of the XMHD model itself. Kinetic studies suggest that Biermann magnetic field generation is strongly reduced for conditions with non-local heat flux \cite{ridgers2021inadequacy}. Our model also needs a more rigorous assessment against a database of various XMHD simulations, preferably including burn and perturbations in three-dimensions. It has only been compared to a single post-processed simulation here. This would help to address several of the caveats and reduce uncertainty in Eq. (\ref{chimodel}). The model could also be extended to treat $L_s$ using the theory of time-dependent Rayleigh-Taylor spikes. Additional physics, such as magnetized alpha transport, should also be studied.

\section*{Authors' contributions}
J.S. conducted the study and wrote the paper. C.W., Y.Z. and H.L. aided with data analysis, discussions and development. 
\begin{acknowledgments}
Research presented in this article was supported by the Laboratory Directed Research and Development program of Los Alamos National Laboratory, under the Center for Nonlinear Studies project number 20190496CR and the Center for Space and Earth Science project number 20180475DR. H.L. acknowledges the support by the Mix and Burn project under the NNSA ASC Physics and Engineering Models program at LANL. The authors thank Brian M. Haines for supply of the xRAGE simulation data previously published in ref. \cite{haines2020cross}.
\end{acknowledgments}
\section*{Data Availability}
The data that support the findings of this study are available from the corresponding author upon reasonable request.

%\bibliography{bibliography.bib}
%\begin{thebibliography}{37}%
%\end{thebibliography}%


\end{document}